\documentclass[conference]{IEEEtran}
\usepackage{amsmath,amssymb,amsfonts,cases,subeqnarray}
\usepackage{graphicx}
\usepackage{textcomp}
\def\BibTeX{{\rm B\kern-.05em{\sc i\kern-.025em b}\kern-.08em
    T\kern-.1667em\lower.7ex\hbox{E}\kern-.125emX}}
\begin{document}

\title{A Two-Ray Multipath Model for Frequency Diverse Array-Based Directional Modulation in MISOME Wiretap Channels}

\author{\IEEEauthorblockN{Qian Cheng$^{\ast\dagger}$, Vincent Fusco$^{\dagger}$, Shilian Wang$^{\ast}$, and Jiang Zhu$^{\ast}$}
\IEEEauthorblockA{
\textit{$^{\ast}$ College of Electronic Science, National University of Defense Technology, Changsha, China} \\
\textit{$^{\dagger}$ Institute of Electronics, Communications and Information Technology (ECIT), Queen's University Belfast, Belfast, UK}\\
chengqian14a@nudt.edu.cn, v.fusco@ecit.qub.ac.uk, \{wangsl, jiangzhu\}@nudt.edu.cn}
}

\maketitle

\begin{abstract}
A two-ray multipath model for frequency diverse array (FDA)-based directional modulation (DM) is proposed in multi-input single-output multi-eavesdropper (MISOME) wiretap channels for the first time. The excitation factors of the FDA and the weighting coefficients of the inserted artificial noise (AN) are jointly designed in a way which imposes no impact on the desired receiver while simultaneously distorting the received  signals of eavesdroppers. Secrecy rate is analyzed for the proposed two-ray multipath FDA-based DM model. Numerical simulations  verify the capability of physical layer secure (PLS) transmissions of the proposed FDA-DM model in two-ray multipath MISOME wiretap channels.
\end{abstract}


\section{Introduction}

Compared with the conventional phased arrays (PA)-based directional modulation (DM) technology \cite{Daly_DM}\cite{Ding_DM}, the frequency diverse arrays (FDA)-based DM scheme is a newly proposed technology with the advantage of wireless physical layer secure (PLS) transmissions not only in  the angle dimension but also in the range dimension.

FDA technology was first proposed in \cite{Antonik_Linear_FDA}, which applied a small frequency increment across the antenna elements and thus produced a range-angle-dependent beampattern. Afterwards, the work in \cite{Wang_Subarray_FDA} divided the FDA elements into multiple subarrays and optimized the transmit beamspace matrix with convex optimization for the purpose of range and angle estimations of targets. An improved FDA radar with logarithmically increasing frequency increments was generalized in \cite{Khan_Log_FDA}, which obtained a beampattern with a single maximum at the target location. Multiple carriers were combined with FDA in \cite{Shao_Dot_FDA} to generate a more focused dot-shaped beampattern. The authors in \cite{Liu_Random_FDA} proposed the random frequency diverse array for uncoupled range-angle indication in active sensing. A focused beampattern synthesis was put forward in \cite{Xiong_GA_FDA} by optimizing the frequency increments with the genetic algorithm, where both single-dot and multidot-shaped transmit beampatterns can be synthesized. Recently, the work in \cite{Basit_Review_FDA} reviewed the developments of the FDA technology.

Owing to the favorable characteristic that the beampattern is both range and angle dependent, FDA technology, in recent years \cite{Xiong_FDA_DM}-\cite{Lin_FDA_DM}, has been applied into DM synthesis for wireless PLS transmissions. Specifically, the authors in \cite{Xiong_FDA_DM} achieved directional modulation synthesis using the FDA with linearly increasing frequency increments. The FDA with symmetrically and non-linearly increasing frequency increments  was used in \cite{Wang_FDA_DM} to decouple range and angle dependent transmit beampattern for wireless DM transmissions. The work in \cite{Hu_Random_FDA_DM} explored the FDA with random frequency increments for DM transmissions. The time-modulated FDA was utilized in \cite{Cheng_Time_FDA_DM} to perform wireless PLS transmissions in free space. Additionally, the work in \cite{Ji_Fading_FDA_DM} showed that FDA-based DM is still achievable in Nakagami-{\emph m} fading channels. The spread spectrum technology was combined with orthogonal FDA in \cite{Xie_OFDA_MB_DM} to achieve multi-beam DM synthesis in free space. Recently, the work in \cite{Lin_FDA_DM} generalized the wireless PLS transmissions based on FDA technology for proximal legitimate user and eavesdropper.

The above-mentioned FDA-based DM schemes \cite{Xiong_FDA_DM}-\cite{Lin_FDA_DM}, however, mainly focused on wireless PLS transmissions in single-path environments without considering multipath effects. Although some works explored the multipath characteristics of the FDA, such as \cite{Cetintepe_Multipath_FDA}, they did not provide any investigations about the possibility of applying the FDA method into DM transmissions in such multipath environments. \emph{To the best of our knowledge, no work is available in the present literature with respect to the capability of wireless PLS transmissions of the FDA-based DM technology in multipath environments. This paper is dedicated, for the first time, to extending the FDA-based DM scheme into multipath environments by proposing a simplified two-ray multipath FDA-DM model in multi-input single-output multi-eavesdropper (MISOME) wiretap channels.}

The rest of this paper is organized as follows. Section II elaborates the main principles of the proposed two-ray multipath FDA-DM model in MISOME wiretap channels. The secrecy rate of the proposed two-ray multipath FDA-DM model is analyzed in Section III. Simulations about bit error rate (BER) and secrecy rate are provided in Section IV. Finally, Section V makes conclusions for the whole paper.

\begin{figure*}
\centering
\includegraphics[angle=0,width=0.65\textwidth]{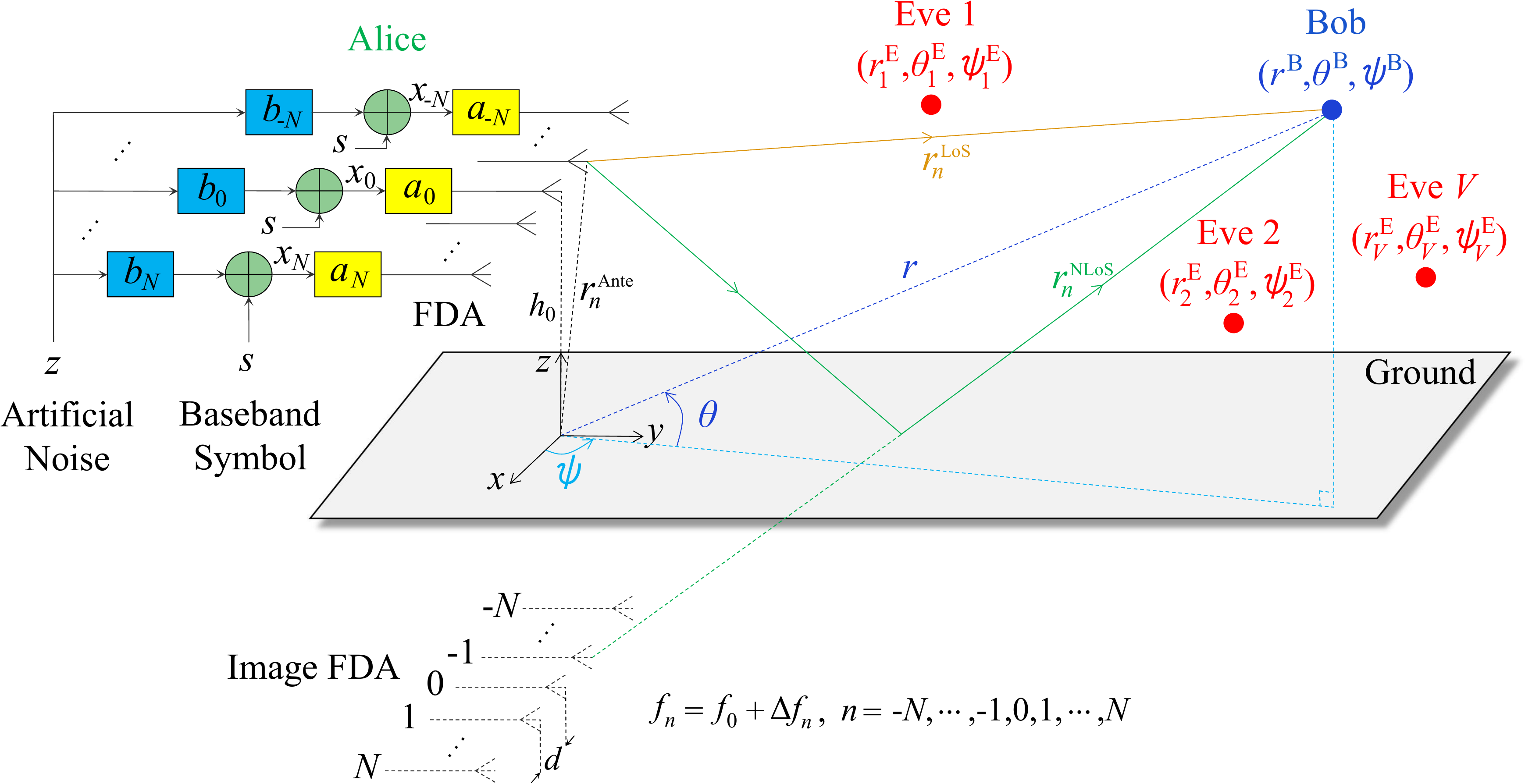}
\caption{The proposed two-ray multipath FDA-based DM model in MISOME wiretap channels.}
\end{figure*}

\emph{Notations}: Throughout the paper,  scalars and vectors are denoted by italic lower-case letters and arrowed bold lower-case letters, respectively. The operator ``$\cdot$'' stands for the product of two vectors. $\mathbb{E}(*)$ means the expectation operation of a random variable, while ${\cal CN}(0,{\sigma ^2})$ refers to the complex Gaussian distribution with zero mean and variance ${\sigma ^2}$. The unit vectors along the $x$, $y$ and $z$-axes of the Cartesian coordinate system are represented by $\vec{\bf{e}}_x$, $\vec{\bf{e}}_y$ and $\vec{\bf{e}}_z$,  respectively. The expression, $\vec{\bf {r}}$ $:$$=$ $(r,\theta,\psi)$,  refers to an arbitrary spatial point/vector $\vec{\bf {r}}$ with coordinates $(r,\theta,\psi)$ in the polar coordinate system. Additionally, $\vec{\bf{e}}_r$ is the unit vector of $\vec{\bf {r}}$.

\section{System Model}

As shown in Fig. 1, we consider an MISOME wiretap model where one transmitter (Alice) with multiple antennas is trying to transmit confidential information to one desired legitimate receiver (Bob), while $V$ anonymous passive eavesdroppers (Eves) are trying to intercept the confidential information.

Alice is composed of a symmetrical ($2N+1$)-element FDA with equal element spacing $d$, which is set as half wavelength of the central carrier. These elements are 1-D linearly arrayed on the $x$-axis with the same height $h_0$ away from the ground. Thus, the  position vector of the $n$-th ($n=-N,\cdots,0,\cdots,N $) element can be expressed as 
\begin{equation}
\label{eq_r_n_ele}
\vec{\bf{r}}^{\rm{Ante}}_n =  d_n  \vec{\bf{e}}_x+ h_0 \vec{\bf{e}}_z
\end{equation} 
where $d_n=nd$ is the $x$-axis coordinate of the $n$-th element.

The radiated frequency of the $n$-th element is designed as 
\begin{equation}
\label{eq_f_n}
f_{n}=f_{0}+\Delta f_{n}=f_{0}+\Delta f\ln (|n|+1)^g
\end{equation}   
where $f_0$ is the central radiated frequency, $\Delta f$ refers to a fixed frequency increment satisfying $|\Delta f| \ll f_{0}$, and $g$ represents an additional factor to control the frequency increments.

Let $\Omega$ denote the alphabet of the normalized baseband modulation symbols with size $|\Omega|=M$. Then the $n$-th antenna element of Alice's FDA is fed with the following signal, 
\begin{equation}
\label{eq_x_n}
x_{n}=\sqrt{P_s}\left (\beta_1 s+\beta_2 b_n z\right)
\end{equation} 
where ${P_s}$ indicates the total transmitting power; $s\in \Omega$  is the normalized baseband modulation symbol with $\mathbb{E}(|s|^2)=1$; $z\sim{\cal CN}(0,1)$ represents the circularly symmetric random complex Gaussian artificial noise (AN); $b_n$ denotes the complex weight coefficient of the AN feeding on the $n$-th element; $\beta_1$ and $\beta_2$ are power splitting factors for the useful signal and the AN, respectively, which satisfy the following constraint,
\begin{equation}
\label{eq_beta}
\beta_{1}^{2}+\beta_{2}^{2}=1
\end{equation}

It is assumed that an arbitrary observation point is located at $\vec{\bf{r}} = r\vec{\bf{e}}_r$ $:$$=$ $(r,\theta,\psi)$, where $r$, $\theta$ and $\psi$ represent the distance from the coordinate origin to the observation point, the  elevation and azimuth angles, respectively.  In such a two-ray multipath model, there exist two transmission paths for the signal transmitting from the $n$-th antenna element to the observation point. The first one is the direct line-of-sight (LoS) path $\vec{\bf{r}}^{\rm{LoS}}_n$, and the second is the reflected non-LoS (NLoS) path  $\vec{\bf{r}}^{\rm{NLoS}}_n$. Geometrically, the length of the direct LoS path $\vec{\bf{r}}^{\rm{LoS}}_n$ can be calculated by \cite{Cetintepe_Multipath_FDA}
\begin{equation}
\label{eq_R_n}
r^{\rm{LoS}}_n \approx r - \vec{\bf{e}}_r\cdot\vec{\bf{r}}^{\rm{Ante}}_n = r-d_n u - h_0 v
\end{equation} 
where $u \triangleq \cos \theta \cos \psi$ and $v \triangleq \sin\theta$.

To facilitate analysis, we suppose that the ground plane is infinite and perfectly conducting, so that the image theory is applicable. Therefore, the NLoS path can be equivalently regarded as an LoS path radiated from the image FDA, as shown in Fig. 1. This means that the length of the NLoS path from the $n$-the element to the observation point can be calculated by
\begin{equation}
\label{eq_R_n2}
r^{\rm{NLoS}}_n \approx r-d_n u + h_0 v
\end{equation}

\newcounter{cnt1}
\setcounter{cnt1}{\value{equation}}
\setcounter{equation}{6}
\begin{figure*}[t]
\begin{subequations}
\label{eq_y_LoS1}
\begin{align}
y^{\rm{LoS}}(\vec{\bf r}) 
&= \sum_{n=-N}^{N}a_n x_n \exp\left\{j2\pi f_n \left(t - \frac{r^{ \rm{LoS}}_n}{c}\right) \right\} = \sum_{n=-N}^{N}a_n x_n \exp\left\{j2\pi f_n \left(t - \frac{r-d_n u - h_0 v}{c}\right) \right\}\\
&= \sum_{n=-N}^{N}a_n x_n \exp\left\{j2\pi \left (f_0+\Delta f_n \right) \left(t - \frac{r - d_n u}{c}\right) \right\}\exp\left\{ j\frac{2\pi f_n h_0 v} {c}\right\}\\
&= \exp\left\{j2\pi f_0 t \right\}\sum_{n=-N}^{N}a_n x_n \exp\left\{j2\pi f_0 \left( - \frac{r - d_n u}{c}\right) \right\}\exp\left\{ j2\pi \Delta f_n \left(t - \frac{r - d_n u}{c}\right)\right\} \exp\left\{ j\frac{2\pi f_n h_0 v} {c}\right\}
\end{align}
\end{subequations} 
\hrulefill
\end{figure*}
\setcounter{equation}{\value{cnt1}}
\setcounter{equation}{7}

\begin{table*}[!t]
\renewcommand{\arraystretch}{1.5}
\caption{One Possible Solution for ${a_n}$ and ${b_n}$ Satisfying (15) with $N=3$}
\label{table1}
\centering
\begin{tabular}{cccccccc}
\hline
 $n$ & $-3$ & $-2$ & $-1$ & $0$ & $1$ & $2$ & $3$\\ [0.1ex]
\hline
$a_n$ & $0.998 + 0.050j$ & $0.516 + 0.857j$ & $0.384 + 0.923j$ & $-0.955 + 0.296j$ & $-0.996 - 0.091j$ & $0.979 - 0.204j$ & $0.222 - 0.975j$\\
$b_n$ & $0.286 - 0.958j$ & $-0.938 + 0.346j$ & $0.535 - 0.845j$ & $-0.949 + 0.315j$ & $-0.995 + 0.105j$ & $-0.971 - 0.238j$ & $-0.336 + 0.942j$\\
\hline
\end{tabular}
\\
\end{table*}

For the sake of convenient analysis, we consider the normalized LoS channel in free space, and ignore the additive white Gaussian noise (AWGN), $\xi\sim{\cal CN}(0,{\sigma_\xi ^2})$. Then the received signal at an arbitrary observation point $\vec{\bf r}$ via the normalized LoS channel can be expressed in (\ref{eq_y_LoS1}), where ${{a}}_n=e^{j{\phi}_n}$ is the excitation factor for the $n$-th element of the FDA.

For clarity, letting
\begin{subequations}
\label{eq_mu_ee}
\begin{numcases}{}
\mu _n = \exp\left\{j2\pi f_0 \left(-\frac{r - d_n u}{c}\right)\right\}\\ 
\varepsilon _n = \exp\left\{ j2\pi \Delta f_n \left(t - \frac{r - d_n u}{c}\right)\right\} 
\end{numcases}{}
\end{subequations}
then the received LoS signal in (\ref{eq_y_LoS1}c) can be written as
\begin{equation}
\label{eq_y_LoS2}
y^{\rm{LoS}}(\vec{\bf r})= \exp\left\{j2\pi f_0 t \right\}\sum_{n=-N}^{N}a_n x_n \mu _n \varepsilon _n \exp\left\{ j\frac{2\pi f_n h_0 v} {c}\right\}
\end{equation}

Similarly, the received signal at the observation point $\vec{\bf r}$ via the normalized NLoS channel is
\begin{equation}
\label{eq_y_NLoS}
\begin{aligned}
&y^{\rm{NLoS}}(\vec{\bf r}) = -\sum_{n=-N}^{N}a_n x_n \exp\left\{j2\pi f_n \left(t - \frac{r^{\rm{NLoS}}_n}{c}\right) \right\}\\
&~~~\approx - \sum_{n=-N}^{N}a_n x_n \exp\left\{j2\pi f_n \left(t - \frac{r-d_n u + h_0 v}{c}\right) \right\}\\
&~~~= - \exp\left\{j2\pi f_0 t \right\}\sum_{n=-N}^{N}a_n x_n \mu _n \varepsilon _n \exp\left\{ -j\frac{2\pi f_n h_0 v} {c}\right\}
\end{aligned}
\end{equation}

\newcounter{cnt2}
\setcounter{cnt2}{\value{equation}}
\setcounter{equation}{10}
\begin{figure*}[t]
\begin{subequations}
\label{eq_y_Total1}
\begin{align}
y^{\rm{Total}}(\vec{\bf r}) 
&= y^{\rm{LoS}}(\vec{\bf r})+y^{\rm{NLoS}}(\vec{\bf r})\\
&= \exp\left\{j2\pi f_0 t \right\}\sum_{n=-N}^{N}a_n x_n \mu _n \varepsilon _n \exp\left\{ j\frac{2\pi f_n h_0 v} {c}\right\} - \exp\left\{j2\pi f_0 t \right\}\sum_{n=-N}^{N}a_n x_n \mu _n \varepsilon _n \exp\left\{ -j\frac{2\pi f_n h_0 v} {c}\right\}\\
&= \exp\left\{j2\pi f_0 t \right\}\sum_{n=-N}^{N} j2 a_n x_n \mu _n \varepsilon _n \sin\left(\frac{2\pi f_n h_0 v} {c}\right) \overset{\text{LP}}{=} \sum_{n=-N}^{N} j2 a_n x_n \mu _n \varepsilon _n \sin\left(\frac{2\pi f_n h_0 v} {c}\right)
\end{align}
\end{subequations} 
\end{figure*}
\setcounter{equation}{\value{cnt2}}
\setcounter{equation}{11}

The total received signal at the observation point $\vec{\bf r}$ can be obtained by adding the LoS signal in (\ref{eq_y_LoS2}) and the NLoS signal in (\ref{eq_y_NLoS}), which is expressed in (\ref{eq_y_Total1}). For clarity, we let
\begin{equation}
\label{eq_rho}
\rho _n = j2\sin\left(\frac{2\pi f_n h_0 v} {c}\right)
\end{equation}

\newcounter{cnt3}
\setcounter{cnt3}{\value{equation}}
\setcounter{equation}{12}
\begin{figure*}[t]
\begin{subequations}
\label{eq_y_Total2}
\begin{align}
y^{\rm{Total}}(\vec{\bf r}) 
 &= \sum_{n=-N}^{N}  a_n x_n \mu _n \varepsilon _n \rho _n = \sum_{n=-N}^{N} a_n \sqrt{P_s}\left(\beta_1 s + \beta_2 b_n z\right)\mu _n \varepsilon _n \rho _n\\
 &= \underbrace {\sqrt{P_s}\beta_1 s \sum_{n=-N}^{N} a_n  \mu _n \varepsilon _n \rho _n}_{\rm{Useful~Signal}} + \underbrace { \sqrt{P_s}\beta_2 z \sum_{n=-N}^{N} a_n  b_n  \mu _n \varepsilon _n \rho _n}_{\rm{Artificial~Noise}} \triangleq \underbrace {\sqrt{P_s}\beta_1  \kappa s }_{\rm{Useful~Signal}} + \underbrace { \sqrt{P_s}\beta_2  \eta z }_{\rm{Artificial~Noise}} 
\end{align}
\end{subequations}
\hrulefill
\end{figure*}
\setcounter{equation}{\value{cnt3}}
\setcounter{equation}{13}

\noindent{Then the total received signal in (\ref{eq_y_Total1}c) can be further simplified as (\ref{eq_y_Total2}), where}
\begin{subequations}
\label{eq_kappa_eta}
\begin{numcases}{}
\kappa \triangleq \sum_{n=-N}^{N} a_n  \mu _n \varepsilon _n \rho _n\\ 
\eta \triangleq \sum_{n=-N}^{N} a_n  b_n  \mu _n \varepsilon _n \rho _n  
\end{numcases}
\end{subequations}

In order to achieve a secure DM transmission from Alice to Bob, the next work is how to design the excitation factors of the FDA $\{ a_n\}$ and the weight coefficients of the inserted AN $\{ b_n\}$. By doing so, only Bob can acquire the useful information while Eves' received signals are severely distorted, thus providing transmission security between Alice and Bob. We suppose that the location information of Bob, $\vec{\bf{r}}^{\rm B}$ $:$$=$ $(r^{\rm B},\theta^{\rm B},\psi^{\rm B})$, is prior known for Alice, and then the design methodologies for $\{ a_n\}$ and $\{ b_n\}$ are given as
\begin{subequations}
\label{eq_a_n_b_n}
\begin{numcases}{}
\kappa^{\rm B}  = \sum_{n=-N}^{N} {{a}}_n  \mu ^{\rm B}_n \varepsilon ^{\rm B}_n \rho ^{\rm B}_n = 1 \\ 
\eta^{\rm B}  = \sum_{n=-N}^{N} a_n  b_n  \mu ^{\rm B}_n \varepsilon ^{\rm B}_n \rho ^{\rm B}_n = 0 
\end{numcases}
\end{subequations}  
where $\mu ^{\rm B}_n$, $\varepsilon ^{\rm B}_n$, and $ \rho ^{\rm B}_n$  are calculated by substituting $(r,\theta,\psi)=(r^{\rm B},\theta^{\rm B},\psi^{\rm B})$ into (\ref{eq_mu_ee}) and (\ref{eq_rho}).

There are multiple solutions for (\ref{eq_a_n_b_n}), one of which is shown in Table I with $N=3$. If the solution sets of $\{a_n\}$ and $\{b_n\}$ are denoted by $\cal A$ and $\cal B$, respectively, the dynamic DM transmission can be achieved by dynamically and randomly generating $\{a_n\}\in\cal A$ and $\{b_n\}\in\cal B$ at symbol rate level. Once $\{a_n\}$ and $\{b_n\}$ are generated with the constraint of (\ref{eq_a_n_b_n}), the total received signal of Bob can be simplified to
\begin{equation}
\label{eq_y_B_Total}
y^{\rm{Total}}(\vec{\bf r}^{\rm{B}}) =  {\sqrt{P_s}\beta_1  \kappa^{\rm{B}} s } +  { \sqrt{P_s}\beta_2  \eta^{\rm{B}} z } = {\sqrt{P_s}\beta_1 s }
\end{equation}
where only the useful signal is left while the inserted AN is eliminated. Therefore, Bob can easily recover the original confidential information from the received signal.

However, by contrast, the received signals of Eves at different locations from Bob will deteriorate seriously. Actually, on the one hand, the received useful signals of Eves are distorted owning to the non-accomplishment of the normalization property in (\ref{eq_a_n_b_n}a); on the other hand, the received signals of Eves are severely interfered by the AN due to the failure of the orthogonal property in (\ref{eq_a_n_b_n}b). Therefore, Eves cannot intercept the confidential information, which guarantees the security of the confidential transmission from Alice to Bob.

\section{Performance Analysis}

BER and secrecy rate are widely used in DM-related literature to measure the performance of DM systems. The BER of DM systems was fully investigated in \cite{Ding_DM_Metrics}, which also holds for the proposed two-ray multipath FDA-DM model. Here we will analyze the  secrecy rate of the proposed two-ray multipath FDA-DM model.

When the total received signal in (\ref{eq_y_Total2}b) is polluted by the AWGN, $\xi\sim{\cal CN}(0,{\sigma_\xi ^2})$, the signal-to-noise ratio (SNR) $\gamma$ at an arbitrary observation point $\vec{\bf {r}}$ can be calculated by
\begin{equation}
\label{eq_SNR}
\gamma (\vec{\bf {r}})  = \frac{P_s \beta_1^2 |\kappa|^2}{\sigma_{\xi}^{2}}
\end{equation} 

Correspondingly, the signal-to-interference-plus-noise ratio (SINR) $\lambda$ at the observation point $\vec{\bf {r}}$ can be calculated by
\begin{equation}
\label{eq_SINR}
\lambda (\vec{\bf {r}})  = \frac{P_s \beta_1^2 |\kappa|^2}{ P_s \beta_2^2 |\eta|^2 + \sigma_{\xi}^{2}}
\end{equation} 

Therefore, the achievable rate $\zeta$ at an arbitrary observation point $\vec{\bf {r}}$ can be written as
\begin{equation}
\label{eq_rate}
 \zeta (\vec{\bf {r}}) = \log_2\{1+\lambda(\vec{\bf {r}})\}
\end{equation}

Especially, the achievable rate of Bob can be obtained by substituting (\ref{eq_a_n_b_n}) into (\ref{eq_SINR}) and (\ref{eq_rate}), i.e.,
\begin{equation}
\label{eq_rate_Bob}
\zeta(\vec{\bf {r}}^{\rm B})   =  \log_2\left\{1+\frac{P_s \beta_1^2}{\sigma_{\xi}^{2}}\right\}
\end{equation}

We further presume that the $V$ eavesdroppers are located at $\vec{\bf {r}}^{\rm E}_{\nu}$ $:$$=$ $(r_\nu^{\rm E},\theta_\nu^{\rm E},\psi_\nu^{\rm E})$, and they are different from Bob's location, which means $\vec{\bf {r}}^{\rm E}_{\nu}\neq\vec{\bf {r}}^{\rm B}$ for $\forall ~\nu\in \{1,2,\cdots,V\}$. Then, the achievable rate of the $\nu$-th Eve, $\zeta(\vec{\bf {r}}_\nu^{\rm E})$, can be obtained by substituting $\vec{\bf {r}}=\vec{\bf {r}}_{\nu}^{\rm E}$ into (\ref{eq_rate}).

Accordingly, the secrecy rate of the proposed two-ray multipath FDA-DM model, $\zeta_{\rm sec}$, can be calculated by
\begin{equation}
\label{eq_Sec_rate}
\zeta_{\rm sec}   = \left[ \min\limits_{\nu\in\{1,\cdots,V\}} \left\{\zeta(\vec{\bf {r}}^{\rm B}) - \zeta(\vec{\bf {r}}_\nu^{\rm E}) \right\}\right]^{+}
\end{equation}
where $[*]^{+}=\max\{0,*\}$.

\begin{figure*}
\centering
\includegraphics[angle=0,width=0.95\textwidth]{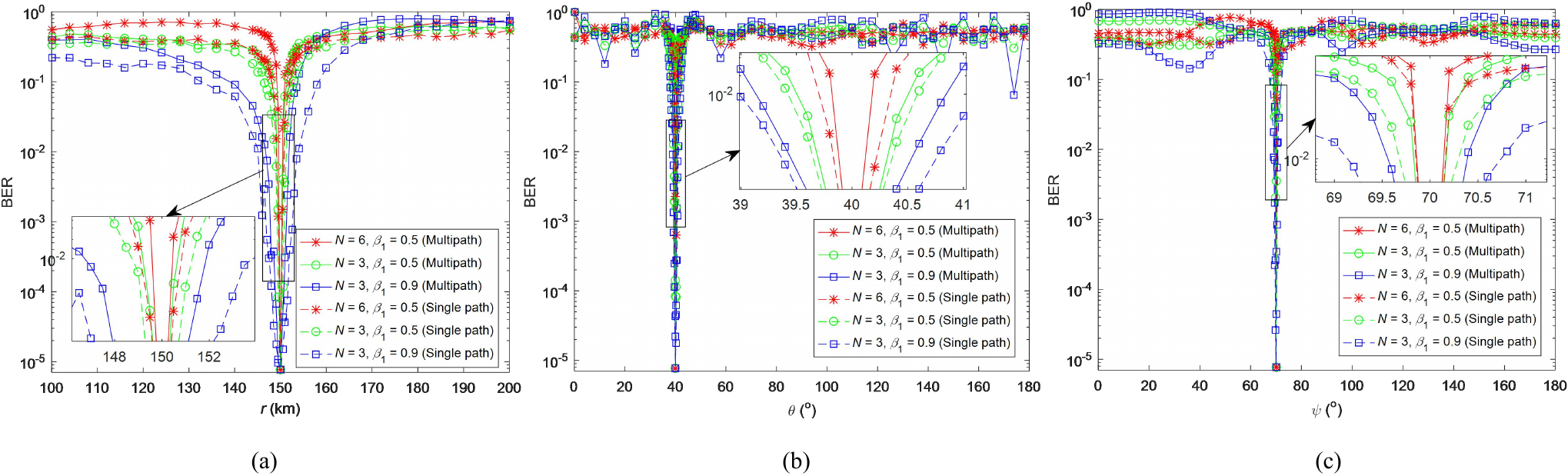}
\caption{BER performances versus (a) range $r$, (b) elevation angle $\theta$, and (c) azimuth angle $\psi$ for the single-path and the proposed multipath FDA-DM models.}
\end{figure*}

\begin{figure*}
\centering
\includegraphics[angle=0,width=0.95\textwidth]{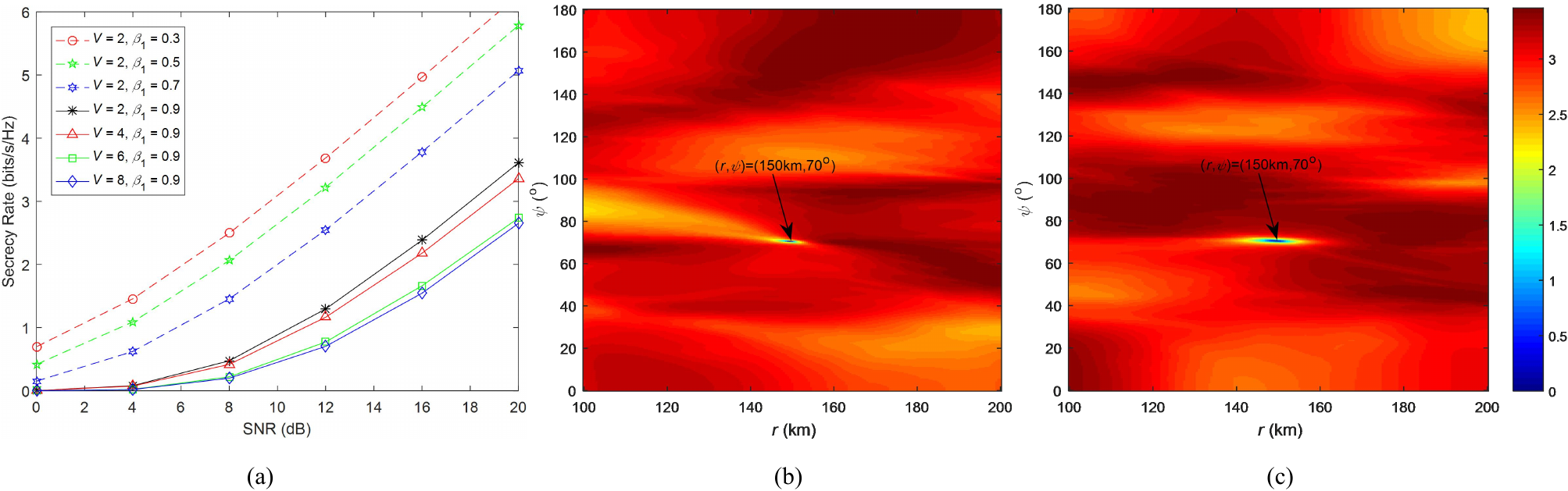}
\caption{Secrecy rate of the proposed multipath FDA-based DM model. (a) Secrecy rate versus SNR (dB); (b) Secrecy rate versus Eve's location (Multipath); (c) Secrecy rate versus Eve's location (Single path).}
\end{figure*}

\section{Simulation Results}

The simulation parameters are selected as $f_0 = 10$ GHz, $\Delta f = 2$ kHz, $g=1$, $(r^{\rm B},\theta^{\rm B},\psi^{\rm B})=(150{\rm km},40^{\circ},70^{\circ})$, $h_0=4.25\lambda_0=4.25c/f_0$, $P_s=1$, $\gamma = 10$ dB, and $\pi/4$-QPSK,  respectively. The locations of the Eves in our simulations are randomly generated.

%

Fig. 2(a)-(c) show the BER performances of the single-path and the proposed multipath FDA-based DM models versus range $r$, elevation angle $\theta$, and azimuth angle $\psi$, respectively. Two observations can be obtained from Fig. 2: a) The BER lobe of the proposed multipath model is narrower than the single-path model, which means that  a more focused secure area can be achieved via the proposed multipath model; b) As expected, larger $N$ and smaller $\beta_1$ can result in narrower BER lobe, which holds  for both single-path and the proposed multipath models. Practically, in order to realize narrower BER selection, the number of array elements and the power splitting factor should be properly chosen via weighing the array complexity and power efficiency.

The secrecy rate of the proposed two-ray multipath FDA-based DM model versus SNR (dB) is depicted in Fig. 3 (a). As expected, it shows that the secrecy rate of the proposed two-ray multipath FDA-based DM model increases with larger SNR, smaller $\beta_1$, and smaller number of Eves. Since the anonymous Eve may locate anywhere in free space, we further simulate the secrecy rates of the multipath and the single-path FDA-DM models versus Eve's location, as shown in Fig. 3 (b) and (c), respectively. Comparing Fig. 3 (b) and (c), it is again verified that the proposed multipath FDA-DM model can achieve more focused secure area than the single-path model.

\section{Conclusion}

We investigated the ability of wireless physical layer secure transmission of an example frequency diverse array-based directional modulation system in two-ray multipath environments. The analytical methodology and numerical simulations verified that the proposed FDA-DM model is capable of wireless PLS transmission in two-ray multipath MISOME wiretap channels, which outperforms the single-path model in terms of more focused secure area.

\end{document}